\documentclass{Interspeech}

\title{Zero-Shot Text-to-Speech as Golden Speech Generator: A Systematic Framework and its Applicability in Automatic Pronunciation Assessment}

\author[affiliation={1}]{Tien-Hong}{Lo}
\author[affiliation={1}]{Meng-Ting}{Tsai}
\author[affiliation={2}]{Yao-Ting}{Sung}
\author[affiliation={2}]{Berlin}{Chen}

\affiliation{Department of Computer Science and Information Engineering}{National Taiwan Normal University}{Taiwan}
\affiliation{Department of Educational Psychology and Counseling}{National Taiwan Normal University}{Taiwan}
\email{\{teinhonglo, mengting7tw, sungtc, berlin\}@ntnu.edu.tw}

\usepackage{comment}
\usepackage{multirow}
\usepackage{graphicx}
\usepackage{booktabs}
\usepackage{subcaption} 

\begin{document}

\maketitle

\begin{abstract}
    
Second language learners can improve their pronunciation by imitating golden speech, especially when the speech that aligns with their respective speech characteristics. This study explores the hypothesis that learner-specific golden speech generated with zero-shot text-to-speech (ZS-TTS) techniques can be harnessed as an effective metric for measuring the pronunciation proficiency of L2 learners. Building on this exploration, the contributions of this study are at least two-fold: 1) design and development of a systematic framework for assessing the ability of a synthesis model to generate golden speech, and 2) in-depth investigations of the effectiveness of using golden speech in automatic pronunciation assessment (APA). Comprehensive experiments conducted on the L2-ARCTIC and Speechocean762 benchmark datasets suggest that our proposed modeling can yield significant performance improvements with respect to various assessment metrics in relation to some prior arts. To our knowledge, this study is the first to explore the role of golden speech in both ZS-TTS and APA, offering a promising regime for computer-assisted pronunciation training (CAPT).

\end{abstract}

\section{Introduction}

With the unprecedented advancements in computer technology and the growing number of second-language (L2) learners worldwide, computer-assisted pronunciation training (CAPT) has emerged as a handy tool for L2 learners. In addition, CAPT can also facilitate instructional quality meanwhile reducing teacher workload. Typically, a CAPT system is deployed in a “read-aloud” scenario where a text prompt is presented, and an L2 learner is asked to say it out loud while mimicking the manner of speech of native speakers. 

\begin{figure}[!t]
    \centering
    \includegraphics[width=\linewidth]{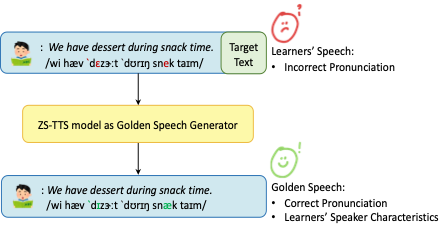}
    \caption{A running example illustrates the advantage of leveraging golden speech for pronunciation training.}
    \label{fig:golden_speech_eg}
\end{figure}

Previous research indicated that L2 learners can significantly improve their pronunciation by imitating a “golden speech” which has a correct pronunciation and speech traits similar to their own \cite{TeachingWatson1989_tvr}, as opposed to imitating those of other native speakers with dissimilar characteristics \cite{GldJilka1998_still,GldNagano1990_icslp,GldProbst2002_spcomm,GldDing2019_spcomm}, as schematically depicted in Figure \ref{fig:golden_speech_eg}. Drawing the benefit of golden speech, we aim to generate learner-specific golden speech using modern synthesis techniques and in turn utilize it for the CAPT-related tasks.

The recent trend of speech synthesis has shifted the focus to drawing on deep neural networks due to their significant technological breakthroughs \cite{TacotronWang2017_interspeech,FastspeechRen2019_nips,VitsKim2021_pmlr,YrttsCasanova2022_pmlr,ZsTTSSaeki2023_ijcai,ZmmttsGong2024_taslp,NaturalSpeechTan2024_tpami}. The prevailing models, including but not limited to Tacotron \cite{TacotronWang2017_interspeech}, FastSpeech \cite{FastspeechRen2019_nips}, VITS \cite{VitsKim2021_pmlr}, and YourTTS \cite{YrttsCasanova2022_pmlr}, have demonstrated marked performance, paving the way for neural network-based speech synthesis. These approaches have revealed several advantages, including high speech quality in terms of intelligibility and naturalness, alongside reduced reliance on handcrafted engineering. In addition, considerable research has been concentrated on few-shot and zero-shot text-to-speech (ZS-TTS) to address the challenge of both insufficient training data and versatile user scenarios \cite{FastspeechRen2019_nips,YrttsCasanova2022_pmlr,ZsTTSSaeki2023_ijcai,ZmmttsGong2024_taslp,NaturalSpeechTan2024_tpami}.

On a separate front, in CAPT systems, automatic pronunciation assessment (APA) is the most indispensable component. APA aims to evaluate the pronunciation proficiency of L2 learners by considering both segmental and suprasegmental features \cite{ProClsFerrer2015_spcomm,ProsodyCaptLi2017_csl,Apa3mFu2022_apsipa,ApaHipamaDo2023_icassp}. Earlier studies on the task of APA only modeled fluency, lexical stress, and intonation independently \cite{ProClsFerrer2015_spcomm,ProsodyCaptLi2017_csl}, while recent research activities explore multi-aspect and multi-granularity modeling \cite{Apa3mFu2022_apsipa,ApaHipamaDo2023_icassp} in order to offer in-depth feedback on the pronunciation quality of an L2 Learner. While the prior arts have shown promising performance in various aspects of prediction, the potential of using the golden speech to enhance the APA models has so far been largely under-explored in the literature.

Building on the above observations, we hypothesize that golden speech can be generated by zero-shot text-to-speech (ZS-TTS) synthesis and can serve as an effective metric for gauging the pronunciation proficiency of L2 learners\footnote{Golden speech is defined as speech that learners find easy to imitate, typically resembling their self-perceived voice. In this work, we adopt a simplified usage of the term to refer to synthesized speech that combines native-like pronunciation with learner-specific traits.}. To validate the above hypothesis, we first utilized word error rate reduction (WERR) from an off-the-shelf ASR model \cite{AsrWhiperXbain2023_interspeech} to measure intelligibility improvements. Additionally, we employ speaker embedding cosine similarity (SECS) \cite{SpkembGE2EWan2018_icassp} and apply the mean opinion score (MOS) \cite{MosUtmosSaeki2022_interspeech} to verify the quality of synthesized speech. We further adopt dynamic time warping (DTW) \cite{DtwMiodonska2016_cbm} to analyze the discrepancies between the original speech and the synthesized speech of L2 learners, offering an alternative measure for pronunciation proficiency. Finally, we validate the effectiveness of golden speech features on a cutting-edge APA model (e.g., 3M \cite{Apa3mFu2022_apsipa}). All variants of our approaches are evaluated on the L2-ARCTIC \cite{CorpusL2arcticZhao2018_interspeech} and the Speechocean762 \cite{CorpusSo762Zhang2021_interspeech} benchmark datasets, which underscore the potential of TTS in generating golden speech for use in personalized CAPT. In summary, the main contributions of this study are as follows:

\begin{enumerate}
    \item We proposed a systematic framework for assessing the ability of a synthesis model to imitate a golden speaker. 
    \item We validated the effectiveness of golden speech features on the APA task and the potential of such fusion to drive further innovative progress in APA. 
    \item In particular, our work is first to explore the role of golden speech in both ZS-TTS and APA, offering a promising regime for CAPT.
\end{enumerate}

\section{Method}

In this section, we delved into the design and implementation of our proposed framework, which unfolded in two parts: 1) the ability of synthesis models to mimic golden speech and 2) assessed the effectiveness of golden speech on APA.

\subsection{Ability of Synthesis Models to Mimic Golden Speech}

\subsubsection{Intelligibility}

Computer-assisted pronunciation training (CAPT) is typically categorized into two types: read-aloud and free-speaking assessments. The former primarily relies on the word error rate (WER) for intelligibility measurement, while the latter also considers grammar and language use. In this study, we focus exclusively on read-aloud assessments and leverage an off-the-shelf automatic speech recognition (ASR) system for our purpose. More specifically, we first utilize the Whisper \cite{AsrWhiperXbain2023_interspeech} model to transcribe both synthesized and original speech into word sequences. We then calculate the word error rate (WER) and the word error rate reduction (WERR) between these sequences.

\subsubsection{Speaker Similarity}

Drawing on prior studies \cite{SpkembGE2EWan2018_icassp}, we calculate the pairwise cosine similarity between original and synthesized speech. After obtaining cosine similarity values, we design two metrics: $SECS_{utt}$, which requires identical linguistic content in the speech, and $SECS_{spk}$, which imposes no such constraint:
\begin{align}
\label{eq:sec_utt}
SECS_{utt}=\frac{1}{S}\sum_{s=1}^S \frac{1}{U_s}\sum_{i=1}^{U_s}cos(\mathbf{e}_i^S, \hat{\mathbf{e}}_i^S )
\end{align}

\begin{align}
\label{eq:sec_spk}
SECS_{spk}=\frac{1}{s}\sum_{s=1}^S\frac{1}{U_S^2}\sum_{i=1}^{U_s}\sum_{j=1}^{U_s}cos(\mathbf{e}_i^S,\hat{\mathbf{e}}_j^S)
\end{align}

In Eq. (\ref{eq:sec_utt}) and Eq. (\ref{eq:sec_spk}),  $U_s$ represents the total number of utterances for speaker $s$, $\mathbf{e}_i^S$ is the speaker embedding of the i-th original utterance for speaker $s$, S signifies the total number of speakers. In Eq. (\ref{eq:sec_utt}), $\hat{\mathbf{e}}_i^S$ corresponds to the speaker embedding of the $i$-th synthesized utterance, reflecting a paired comparison between the original speech and the synthesized speech. In contrast, Eq. (\ref{eq:sec_spk}) allows for comparisons across different utterances for the same speaker to evaluate the speaker similarity value. The only distinction between the two equations lies in the comparison for the same (i $=$ j in Eq. (\ref{eq:sec_utt})) or across different utterances (i $\neq$ j in Eq. (\ref{eq:sec_spk})).

\subsubsection{Naturalness}

The mean opinion score (MOS) is a widely-used metric for the evaluation of speech naturalness, relying on subjective scores from human listeners. Recently, the development of automatic MOS models, which are built on deep neural networks to predict MOS values, can reduce costs and improve the reproducibility of synthesis models. We adopt SpeechMOS , an automatic MOS model derived from UTMOS \cite{MosUtmosSaeki2022_interspeech}, to assess naturalness. SpeechMOS has shown a high correlation with MOS scores provided by human annotators, thereby suggesting it as a viable alternative for subjective evaluations.

\subsection{Effectiveness of Golden Speech on APA}
\subsubsection{DTW-based Correlation}

We assume that the smaller the difference between the original speech and the synthesized (golden) speech of an L2 learner, the better the pronunciation proficiency of the learner. To validate this assumption, we first employ dynamic time warping (DTW) to calculate the differences (i.e., DTW cost) between the original speech and the synthesized speech based on their respective feature vector sequences (e.g., wav2vec 2.0 \cite{SslLarochelle2020_nips}). We then assess the correlation between the DTW costs and the pronunciation proficiency scores (e.g., TOEFL and total score) of L2 learners. A correlation value between the DTW costs and the proficiency score of the L2 learner is anticipated to indicate the effectiveness of using learner-specific golden speech as an important ingredient for APA.

\begin{figure}[!t]
    \centering
    \includegraphics[width=\linewidth]{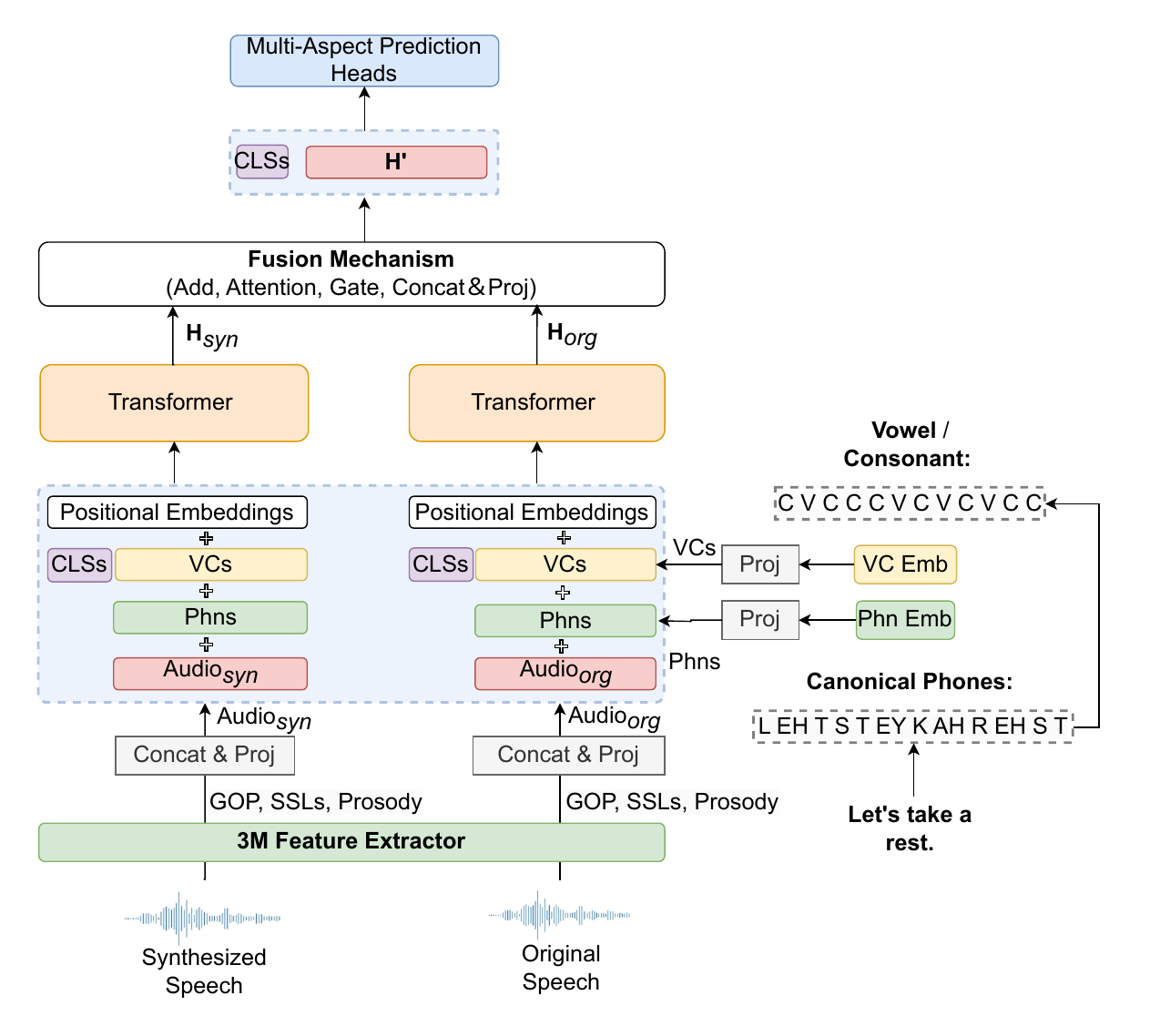}
    \caption{A schematic diagram of our proposed golden speech modeling for APA. This architecture is implemented in our 3M-based APA model and evaluated in Section 3.6, where we test its effectiveness under different fusion strategies.}
    \label{fig:3m_3mtts_arch}
\end{figure}

\subsubsection{Golden Speech-based APA Model}
This work selects the 3M model \cite{Apa3mFu2022_apsipa}, a cutting-edge APA model, as the foundation model to facilitate the validation of our hypothesis. 3M extracts multi-view inputs of speech of learners and delivers feedback across multiple aspects (e.g., accuracy, total, etc.) and granularities (e.g., phoneme, word, and utterance) simultaneously. Building upon the 3M model, we further enhance its performance by utilizing the synthesized (golden) speech of the L2 learner. Our proposed architecture, as depicted in Figure \ref{fig:3m_3mtts_arch}, includes a synthesized speech transformer encoder that mirrors the original 3M transformer and incorporates four fusion mechanisms: 1) \textbf{Addition (ADD)}: The addition mechanism merges the hidden vector sequences of the original speech, $\mathbf{H}_{org}$, and the synthesized speech, $\mathbf{H}_{syn}$, by summing them together. 2) \textbf{Attention (ATT)}: In the attention mechanism, we designate $\mathbf{H}_{org}$ as the query, with $\mathbf{H}_{syn}$ serving both as the key and the value. $\mathbf{H}^{\prime}$ is derived using multi-head attention MHA and subsequently added to $\mathbf{H}_{org}$ in a residual manner. 3) \textbf{GATE}: In the gate mechanism, the weight matrices $\mathbf{W}_{org}$ and $\mathbf{W}_{syn}$ are introduced to specify the importance of the original speech $\mathbf{H}_{org}$ and the synthesized speech $\mathbf{H}_{syn}$, respectively. 4) \textbf{Concatenation (CAT)}: In the CAT mechanism, $\mathbf{H}_{org}$ and $\mathbf{H}_{syn}$ are concatenated and followed by MLP layer Linear to obtain the $\mathbf{H}^{\prime}$. The fused hidden representation $\mathbf{H}^\prime$ is passed to the multi-aspect prediction heads for APA scoring.

\section{Experiments}
\subsection{Dataset}
\textbf{\textit{L2-ARCTIC}} is an English speech benchmark dataset designed and curated for research in voice conversion, accent conversion, and mispronunciation detection and diagnosis, among others. This corpus contains about 3,600 utterances of 24 non-native speakers (12 males and 12 females, equipped with manual transcripts) across different nationalities. The dataset features about one hour of recorded speech per speaker, including challenging sentences for the L1 of non-native speakers. Furthermore, it provides manual phoneme-level annotations for mispronunciations in 150 sentences. In this study, L2-ARCTIC serves as an evaluation set for quantitative analysis of the synthesized golden speech for L2 learners. 

\textbf{\textit{Speechocean762}} is an open-source dataset designed for pronunciation assessment, containing 5,000 English utterances from 250 non-native speakers, ranging from adults (over 15 years old) to children (15 years old or younger). The Speechocean762 (SO762) dataset includes scores for pronunciation proficiency across multiple dimensions and at various levels of linguistic granularity, equipped with phoneme-level, word-level, and utterance-level annotation labels (i.e., pronunciation proficiency scores) with respect to different aspects, such as accuracy, fluency, prosody, and a total score. We selected the accuracy and total scores at phoneme, word, and utterance levels for evaluation to facilitate the validation of our hypothesis.

\subsection{Implementation Details}

In our experiments, we validated the proposed approach along two directions. The first focused on the synthesis model’s ability to mimic golden speech (\textit{cf}. Section 2.1). We employed the multilingual YourTTS model from Coqui-ai/TTS \cite{YrttsCasanova2022_pmlr} to generate synthesized speech and assessed its intelligibility using WhisperX \cite{AsrWhiperXbain2023_interspeech}, with medium-sized monolingual models (\textit{cf}. Section 2.2.1). Speaker similarity was measured via cosine similarity of speaker embeddings using Resemblyzer (\textit{cf}. Section 2.2.2), and naturalness was evaluated through Mean Opinion Score (MOS) (\textit{cf}. Section 2.2.3).
The second direction evaluated the effectiveness of golden speech in automatic pronunciation assessment (APA) (\textit{cf}. Section 2.2). We used DTW alignment from the tslearn toolkit to compare the original and synthesized speech based on wav2vec 2.0-base feature vectors \cite{SslLarochelle2020_nips}. The synthesized speech was then integrated as an auxiliary input into the APA model 3M \cite{Apa3mFu2022_apsipa}. Following prior work \cite{ApaHipamaDo2023_icassp,Apa3mFu2022_apsipa}, we ran each experiment five times with different seeds and reported the mean and standard deviation of the Pearson Correlation Coefficient (PCC). Code is available at \url{https://anonymous.4open.science/r/l2arctic_tts-F600/}.

\begin{table}[!t]
\centering
\caption{Performance of proposed metrics (i.e., intelligibility, speaker similarity, naturalness) evaluated on both L1 (a speaker from the YourTTS training set) and golden speech (GLD) using text from the L2-ARCTIC (L2Arc) dataset, as well as adult (SO762-A) and children (SO762-C) speech from the Speechocean762 dataset.}
\label{tab:exp_wer_secs_mos_results}
\begin{tabular}{lclll}
\hline
\multirow{2}{*}{} & \multicolumn{1}{l}{\multirow{2}{*}{WER (WERR\%)}} & \multicolumn{2}{l}{SECS} & \multirow{2}{*}{MOS} \\ \cline{3-4}
                 & \multicolumn{1}{l}{} & utt  & spk  &      \\ \hline
\textbf{L2-Arc}  & 7.42 (-)             & 1.00 & 0.83 & 3.86 \\
- L1             & 4.84 (34.77)         & 0.51 & 0.50 & 3.50 \\
-- GLD           & 5.03 (32.21)         & 0.75 & 0.70 & 3.67 \\ \hline
\textbf{SO762-A} & 21.07 (-)            & 1.00 & 0.82 & 3.21 \\
- L1             & 4.32 (79.50)         & 0.46 & 0.45 & 3.52 \\
- GLD            & 4.63 (78.03)         & 0.67 & 0.64 & 3.47 \\ \hline
\textbf{SO762-C} & 25.02 (-)            & 1.00 & 0.77 & 2.93 \\
- L1             & 15.95 (36.25)        & 0.44 & 0.43 & 3.47 \\
- GLD            & 16.47 (34.17)        & 0.64 & 0.59 & 3.31 \\ \hline
\end{tabular}
\end{table}

\subsection{Intelligibility}

At the outset, we evaluate here the capacity of YourTTS to generate golden speech using speech datasets of L2 speakers from Speechocean762 (SO762), comprising both adults (referred to as SO762-A) and children (referred to as SO762-C), as well as from L2-ARCTIC (referred to as L2-Arc). Table \ref{tab:exp_wer_secs_mos_results} shows the corresponding results of the original speech (L2-Arc, SO762-A, SO762-C), the native speech (L1, a seen speaker in the training set of the YourTTS model), and the golden speech (GLD). As can be seen, there is a significant and consistent improvement pertaining to two metrics, i.e., WER and WERR, for synthesized speech (L1 and GLD) across both L2-ARCTIC and SO762 datasets. 
For L2-ARCTIC, the WER for the L1 speech is improved from 7.90\% to 4.63\% with the medium english Whisper model, leading to a WERR of 34.70\%. The golden speech also benefits from the same configuration change of Whisper, with a WERR of 31.59\%. As for SO762-C, a similar trend to L2-ARCTIC is observed, with WERRs of 31.32\% for the L1 speech and 34.03\% for the golden speech. 
These results indicate that while both the L1 speech and the golden speech have better intelligibility, the performance of the latter (GLD) on WERR was not as good as the former (L1). This discrepancy is probably due to the fact that L2 speakers were excluded from the training data, suggesting that including L2 speakers in the training dataset could enhance the performance of ZS-TTS. Nevertheless, the WERR of the golden speech, though slightly lower than that of the L1 speech, is within an acceptable range. We will continue to validate the effectiveness of golden speech (GLD) as a robust feature in subsequent experiments.

\subsection{Speaker Similarity and Naturalness}

Table \ref{tab:exp_wer_secs_mos_results} reports the SECS values (speaker similarity) and CMOS (naturalness). Higher values in both metrics are desirable. We include the original speech here as an upper bound, shown in the first row of each block (L2-Arc, SO762-A, SO762-C). From Table \ref{tab:exp_wer_secs_mos_results}, we can observe the following comparisons between L1 and GLD: For L2Arc, the SECS values for GLD are 0.75 (utt) and 0.70 (spk), compared to 0.51 (utt) and 0.50 (spk) for L1, respectively, showing a higher speaker similarity for GLD. The CMOS for GLD is 3.67, higher than 3.50 for L1. This trend is consistent across SO762-A and SO762-C, in comparison to lower SECS values for L1 in both cases. In SO762, GLD achieves better SECS in the adult group (SO762-A) compared to the children group (SO762-C), due likely to a closer match between the speaker characteristics of adults and the training data. While CMOS does not show significant correlations, satisfactory values are still reached for L2Arc, SO762-A, and SO762-C, respectively. These scores are comparable to those of the original and L1 speech, highlighting that the golden speech retains naturalness while improving intelligibility. Table \ref{tab:exp_wer_secs_mos_results} also reveals that GLD has higher speaker similarity compared to L1, retaining more of the original voicing characteristics of L2 speakers. Next, we will turn to explore the effectiveness of golden speech (GLD) in APA.

\begin{figure}[!t]
    \centering
    \subfloat[]{
        \includegraphics[width=\linewidth]{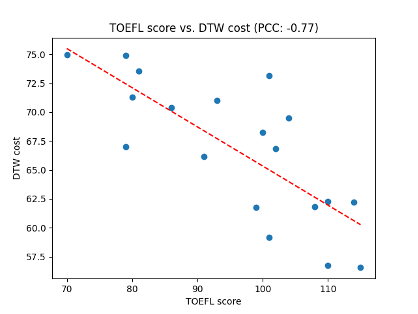}%
        \label{fig:dtw_toefl}
    }
    \vfill
    \subfloat[]{
        \includegraphics[width=\linewidth]{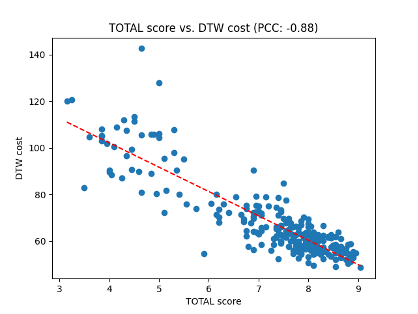}%
        \label{fig:dtw_total}
    }
    \caption{Distribution of DTW cost and pronunciation proficiency score on wav2vec 2.0 embedding. We evaluated the TOEFL score of L2-ARCTIC (\ref{fig:dtw_toefl}) and the total score of Speechocean762 (\ref{fig:dtw_total}), respectively.}
    \label{fig:boxplot_dtw}
\end{figure}

\subsection{DTW cost}

To prove the effectiveness of golden speech in APA, we examine the correlation between DTW costs, calculated using original speech and the golden speech with the wav2vec 2.0 embeddings, as well as the pronunciation proficiency, indicated by TOEFL scores for L2-ARCTIC and total scores for SO762. As shown in Figure \ref{fig:boxplot_dtw}, the PCC between DTW cost and TOEFL scores is -0.77 for L2-ARCTIC and -0.88 for SO762. This demonstrates a clear negative trend, reinforcing the strong correlation between the DTW cost and the proficiency scores. The larger number of speakers in SO762 also makes it more representative. These findings confirm that DTW cost is an effective metric for assessing pronunciation proficiency, with lower DTW costs correlating with higher proficiency scores.

\begin{table}[!t]
\centering
\caption{Performance of different fusion mechanisms evaluated on the Speechocean762 dataset.}
\label{tab:exp_3mtts}
\begin{tabular}{l|c|cc|c}
\hline
\textbf{} &
  \textbf{Phone} &
  \multicolumn{2}{c|}{\textbf{Word}} &
  \textbf{Utt} \\
\textbf{} &
  \textbf{Acc.} &
  \textbf{Acc.} &
  \textbf{Total} &
  \textbf{Total} \\ \hline
\textbf{HiPaMA} &
  \begin{tabular}[c]{@{}c@{}}0.616\\ ±0.004\end{tabular} &
  \begin{tabular}[c]{@{}c@{}}0.575\\ ± 0.004\end{tabular} &
  \begin{tabular}[c]{@{}c@{}}0.591\\ ±0.004\end{tabular} &
  \begin{tabular}[c]{@{}c@{}}0.754\\ ± 0.002\end{tabular} \\
\textbf{3M} &
  \begin{tabular}[c]{@{}c@{}}0.636\\ ±0.002\end{tabular} &
  \begin{tabular}[c]{@{}c@{}}0.598\\ ± 0.003\end{tabular} &
  \begin{tabular}[c]{@{}c@{}}0.614\\ ± 0.004\end{tabular} &
  \begin{tabular}[c]{@{}c@{}}0.798\\ ±0.006\end{tabular} \\ \hline
\textbf{+ADD} &
  \begin{tabular}[c]{@{}c@{}}0.639\\ ± 0.003\end{tabular} &
  \begin{tabular}[c]{@{}c@{}}0.608\\ ± 0.009\end{tabular} &
  \begin{tabular}[c]{@{}c@{}}0.624\\ ± 0.009\end{tabular} &
  \begin{tabular}[c]{@{}c@{}}0.801\\ ± 0.003\end{tabular} \\
\textbf{+ATT} &
  \begin{tabular}[c]{@{}c@{}}0.629\\ ± 0.009\end{tabular} &
  \begin{tabular}[c]{@{}c@{}}0.605\\ ± 0.006\end{tabular} &
  \begin{tabular}[c]{@{}c@{}}0.620\\ ± 0.007\end{tabular} &
  \begin{tabular}[c]{@{}c@{}}0.799\\ ± 0.002\end{tabular} \\
\textbf{+GATE} &
  \textbf{\begin{tabular}[c]{@{}c@{}}0.647\\ ± 0.013\end{tabular}} &
  \textbf{\begin{tabular}[c]{@{}c@{}}0.615\\ ± 0.010\end{tabular}} &
  \textbf{\begin{tabular}[c]{@{}c@{}}0.628\\ ± 0.010\end{tabular}} &
  \begin{tabular}[c]{@{}c@{}}0.804\\ ± 0.004\end{tabular} \\
\textbf{+CAT} &
  \begin{tabular}[c]{@{}c@{}}0.630\\ ± 0.007\end{tabular} &
  \begin{tabular}[c]{@{}c@{}}0.603\\ ±0.005\end{tabular} &
  \begin{tabular}[c]{@{}c@{}}0.616\\ ±0.006\end{tabular} &
  \textbf{\begin{tabular}[c]{@{}c@{}}0.806\\ ± 0.011\end{tabular}} \\ \hline
\end{tabular}
\end{table}

\subsection{Performance on the APA Models}

In this subsection, we report on the performance of two strong baselines, which are multi-granularity multi-aspect APA models, i.e., HiPAMA \cite{ApaHipamaDo2023_icassp} and 3M \cite{Apa3mFu2022_apsipa}. Table \ref{tab:exp_3mtts} shows that our proposed approaches can yield small yet consistent improvements. These results underscore the vital role of golden speech traits in accurately evaluating multi-aspect pronunciation proficiency. The observed limitations for further improvement may stem from using grapheme-to-phoneme (G2P) conversion to generate phoneme sequences for the input of YourTTS, rather than canonical phoneme sequences from human annotations. This discrepancy could have constrained the effectiveness of the golden speech for the phone-level APA.

\section{Conclusion}

In conclusion, we have presented a novel and systematic framework for evaluating the potential of ZS-TTS as a golden speech generator and incorporating the golden speech as an indicative feature into automatic pronunciation assessment (APA). Experiments conducted on the L2-ARCTIC and Speechocean762 datasets have shown notable improvements with respect to WER and other evaluation metrics, confirming that the golden speech generated by ZS-TTS plays an important role in APA. Future studies will explore more sophisticated ZS-TTS architectures, as well as joint training and fine-tuning strategies to further unravel the potential role of the golden speech in APA and related applications \cite{AsaLo2024_naacl}.

\bibliographystyle{IEEEtran}
\bibliography{mybib}

\end{document}